%% file: pbar.tex
\documentclass[aps,prd,showpacs,floatfix,nofootinbib,twocolumn,10pt]{revtex4}

\usepackage{color,amsmath,amssymb,graphicx,latexsym,subfigure}
\usepackage{threeparttable,txfonts}

\def\nat{Nature\ }
\def\aap{Astron.\ Astrophys.\ }
\def\apj{Astrophys.\ J.\ }
\def\apjl{Astrophys.\ J.\ Lett.\ }
\def\apjs{Astrophys.\ J.\ Supp.\ }
\def\aj{Astron.\ J.\ }
\def\mnras{Mon.\ Not.\ Roy.\ Astron.\ Soc.\ }
\def\physrep{Phys.\ Rept.\ }
\def\prd{Phys.\ Rev.\ D\ }
\def\prl{Phys.\ Rev.\ Lett.\ }
\def\apss{Astrophys.\ Space\ Sci.\ }
\def\araa{Annu.\ Rev.\ Astron.\ Astrophys.\ }
\def\jcap{J.\ Cosmol.\ Astropart.\ Phys.\ }

\def\sv{\ensuremath{\langle\sigma v\rangle}}

\begin{document}

\title{Possible dark matter annihilation signal in the AMS-02 antiproton data}

\author{Ming-Yang Cui$^{a,b}$}
\author{Qiang Yuan$^{a,c}$\footnote{The corresponding author: yuanq@pmo.ac.cn}}
\author{Yue-Lin Sming Tsai$^{b,d}$}
\author{Yi-Zhong Fan$^{a,c}$\footnote{The corresponding author: yzfan@pmo.ac.cn}}

\affiliation{
$^a$Key Laboratory of Dark Matter and Space Astronomy, Purple
Mountain Observatory, Chinese Academy of Sciences, Nanjing
210008, P.R.China \\
$^b$Department of Physics, Nanjing University, Nanjing 210093, P.R.China \\
$^c$School of Astronomy and Space Science, University of Science and Technology of China, Hefei, Anhui 230026, P.R.China\\
$^d$Physics Division, National Center for Theoretical Sciences, Hsinchu, 
Taiwan
}

\begin{abstract}

Using the latest AMS-02 cosmic ray antiproton flux data, we search
for potential dark matter annihilation signal. The background parameters
about the propagation, source injection, and solar modulation are not
assumed {\it a priori}, but based on the results inferred from the recent
B/C ratio and proton data measurements instead. The possible dark matter 
signal is incorporated into the model self-consistently under a Bayesian 
framework. Compared with the astrophysical background only hypothesis, 
we find that a dark matter signal is favored. The rest mass of the dark 
matter particles is $\sim 20-80$ GeV and the velocity-averaged 
hadronic annihilation cross section is about $(0.2-5)\times 10^{-26}$ 
cm$^{3}$s$^{-1}$, in agreement with that needed to account for the 
Galactic center GeV excess and/or the weak GeV emission from dwarf 
spheroidal galaxies Reticulum 2 and Tucana III. Tight constraints on 
the dark matter annihilation models are also set in a wide mass region. 

\end{abstract}

\date{\today}


\maketitle

{\it Introduction} ---
The precise measurements of cosmic ray (CR) anti-particle spectra by 
space-borne instruments, such as PAMELA and AMS-02, provide very good 
sensitivity to probe the particle dark matter (DM) annihilation or decay 
in the Milky Way. The CR antiprotons, primarily come from the inelastic 
collisions between the CR protons (and Helium) and the interstellar 
medium (ISM), are effective to constrain the DM models 
\cite{2006JCAP...05..006B,2009PhRvL.102g1301D,2014JCAP...12..045C}. 
Recent observations of the antiproton fluxes \cite{2009PhRvL.102e1101A,
2010PhRvL.105l1101A,2016PhRvL.117i1103A} are largely consistent with the 
expectation from the CR propagation model, leaving very limited room for 
the annihilation or decay of DM \cite{2009PhRvL.102g1301D,
2014JCAP...04..003F,2015JCAP...09..023G,2015PhRvD..92e5027J,
2015arXiv150407230L}.

There are several sources of uncertainties in using antiprotons to 
constrain DM models. The largest uncertainty may come from the propagation 
parameters. Usually the secondary-to-primary ratio of CR nuclei, such
as the Boron-to-Carbon ratio (B/C), and the radioactive-to-stable 
isotope ratio of secondary nuclei, such as the Beryllium isotope ratio
$^{10}$Be/$^9$Be, are used to determine the propagation parameters
\cite{1990cup..book.....G,2007ARNPS..57..285S}. Limited by the data
quality, the constraints on the propagation parameters are loose 
\cite{2001ApJ...555..585M,2010A&A...516A..66P}. Even the
effect on the background antiproton flux due to uncertainties of 
propagation parameters is moderate, the flux from the DM component 
depends sensitively on propagation parameters \cite{2004PhRvD..69f3501D}.
Additional uncertainties include the injection spectrum of the CR nuclei, 
solar modulation, and hadronic interaction models \cite{2015JCAP...09..023G}.
Those uncertainties make the DM searches with antiprotons 
inconclusive \cite{2014PhRvD..90l3001B,2015JCAP...03..021H}.

Given the new measurements of the proton, Helium, and B/C data by 
PAMELA and AMS-02 \cite{2011Sci...332...69A,2015PhRvL.114q1103A,
2015PhRvL.115u1101A,2016PhRvL.117w1102A}, improved constraints on the 
propagation and source injection parameters can be obtained through
global Bayesian approaches \cite{2011ApJ...729..106T,2015JCAP...09..049J,
2016arXiv160706093K,2016ApJ...824...16J}. With these data, we conduct 
a global study to determine the propagation, injection, and solar 
modulation parameters simultaneously using the Markov Chain 
Monte Carlo (MCMC) method \cite{2017arXiv170106149Y}. These ``background'' 
parameters and their likelihoods can be incorporated in the study of 
the DM model parameters by means of the Bayesian theorem, giving 
self-consistent and unbiased judgement of the DM models (see earlier
attempts \cite{2015JCAP...03..021H,2015JCAP...09..049J}). In this work 
we apply this method to the most recently reported antiproton fluxes 
measured by AMS-02 \cite{2016PhRvL.117i1103A}. Furthermore, we improve 
the constraints on the solar modulation parameters with the time-dependent 
proton fluxes measured by PAMELA \cite{2013ApJ...765...91A}.
Note, however, we adopt a relatively simple one-zone diffusion
model in this work. It is possible that in reality the ISM and CR 
propagation are more complicated, e.g., vary everywhere
\cite{2016ApJ...824...16J}.

{\it Background} ---
Here we simply introduce the fitting procedure to determine the propagation, 
injection, and solar modulation parameters \cite{2017arXiv170106149Y,Sup-A}. 
Hereafter they are referred to as background parameters. We work in the 
diffusion reacceleration\footnote{Other propagation scenarios, such as 
the plain diffusion and diffusion convection models, have also been 
discussed in \cite{2017arXiv170106149Y}. However, 
the diffusion reacceleration scenario is found to be best consistent with 
the data.} framework of the CR propagation, which was found to reproduce 
the peak of the B/C data around 1 GeV/n well \cite{2002ApJ...565..280M}. 
The injection spectrum of nuclei is assumed to be a broken power-law with 
respect to rigidity. Although the spectrum of Helium (and heavier nuclei) 
is found to be harder than that of protons, we assume a unified set of 
injection parameters of all nuclei. Such an assumption is expected to not 
sensitively affect the calculation of the B/C ratio. The solar modulation 
model is adopted to be the force-field approximation 
\cite{1968ApJ...154.1011G}. 
As for the modulation potential, we employ a time-variation form 
$\Phi=\Phi_0+\Phi_1\times \tilde{N}(t)$ to connect the modulation with 
solar activities which are characterized by the sunspot 
number\footnote{http://solarscience.msfc.nasa.gov/SunspotCycle.shtml}
$\tilde{N}(t)$ (normalized to 1 at solar maximum of cycle 24). 
The data used in the fitting include the B/C data by ACE 
\cite{2009ApJ...698.1666G} and AMS-02 \cite{2016PhRvL.117w1102A}, the 
proton spectrum by AMS-02 \cite{2015PhRvL.114q1103A} and the time-dependent 
proton fluxes by PAMELA \cite{2013ApJ...765...91A}. The $^{10}$Be/$^9$Be 
ratio is not well measured yet. We use some old data in the fitting
(see Ref. \cite{2017arXiv170106149Y}).

\begin{figure}[!htb]
\includegraphics[width=0.45\textwidth]{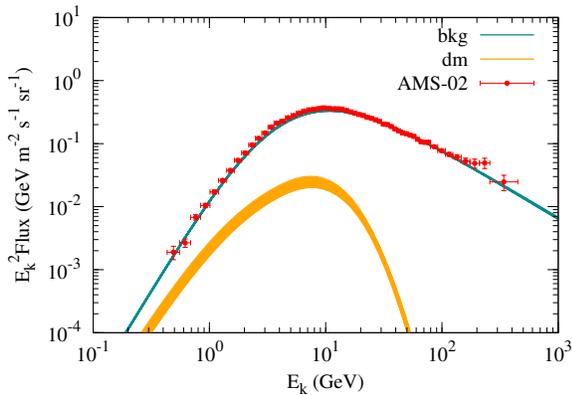}
\caption{Secondary and DM annihilation antiproton fluxes calculated for
$2\sigma$ ranges of the background parameters determined in the fitting
to the B/C, $^{10}$Be/$^9$Be, and proton data. As an illustration, 
the mass of the DM particle is 47 GeV, the cross section is 
$10^{-26}$ cm$^3$s$^{-1}$, and the annihilation channel is $b\bar{b}$.
\label{fig:apflux}}
\end{figure}

The numerical tool GALPROP \cite{1998ApJ...509..212S,1998ApJ...493..694M}
is adopted to calculate the propagation of CRs. We have developed a
global fitting tool, {\tt CosRayMC}, which incorporates GALPROP into the 
MCMC sampler \cite{2002PhRvD..66j3511L}, enabling efficient survey of the 
high-dimensional parameter space of the CR propagation 
\cite{2010PhRvD..81b3516L,2012PhRvD..85d3507L}. Once the background
parameters are obtained, the secondary production of antiprotons can
be obtained, as shown in Fig. \ref{fig:apflux}. Note that there are 
uncertainties from the antiproton production cross section 
\cite{1983JPhG....9.1289T,2003PhRvD..68i4017D,2014PhRvD..90h5017D,
2014JCAP...09..051K,Sup-B}. 
Especially it has been found that an asymmetry exists between the 
antineutron and antiproton production for $pp$ collisions, which tends 
to give more antineutrons \cite{2003APHHI..17..369F}. An energy-independent 
rescaling factor of $\kappa\simeq1.3\pm0.2$ has been suggested to 
approximate the ratio of antineutron-to-antiproton production cross sections 
\cite{2014PhRvD..90h5017D}. The energy-dependence of $\kappa$ is
unclear at present \cite{2014PhRvD..90h5017D,2014JCAP...09..051K}. 
We expect that a constant factor is a simple and reasonable assumption. 
For the results shown in Fig. \ref{fig:apflux} we adopt $\kappa=1.2$.

{\it DM annihilation} ---
Antiprotons can also be produced via the DM annihilation or decay.
In this work we focus on the discussion of DM annihilation. The density
profile of DM is adopted to be NFW profile \cite{1997ApJ...490..493N},
$\rho(r)=\rho_s\left[(r/r_s)(1+r/r_s)^2\right]^{-1}$, where $r_s=20$ kpc
and $\rho_s=0.26$ GeV cm$^{-3}$. The production spectrum of antiprotons
is calculated using the tables given in \cite{2011JCAP...03..051C}. 
Fig. \ref{fig:apflux} shows the results of DM induced antiproton fluxes, 
for $m_{\chi}=47$ GeV and $\sv=10^{-26}$ cm$^3$s$^{-1}$ (for illustration), 
and various background parameters which lie in the $2\sigma$ ranges derived 
in the background fitting. Due to the improved constraints on the propagation
parameters (e.g., the half height of the propagation halo $z_h=5.9\pm1.1$
kpc \cite{Sup-A}), the DM annihilation induced antiproton 
fluxes are constrained in a range of a factor of $\sim2$, which improve 
much compared with previous studies (e.g., \cite{2004PhRvD..69f3501D,
2015JCAP...09..049J}).

{\it Results of DM constraints} ---
From the Bayesian theorem, one can always update the prior from independent 
measurements. The posterior probability density of the parameter $\sv$ for 
given mass of the DM particle $m_{\chi}$ can be written as
\begin{equation}
{\mathcal P}(\sv)|_{m_{\chi}}\propto\int{\mathcal L}(m_{\chi},\sv,
\boldsymbol{\theta}_{\rm bkg},\kappa)\,p(\boldsymbol{\theta}_{\rm bkg})\,
p(\kappa)\,{\rm d}\boldsymbol{\theta}_{\rm bkg}\,{\rm d}\kappa,\nonumber
\end{equation}
where ${\mathcal L}$ is the likelihood function of model parameters 
$(m_{\chi},\sv,\boldsymbol{\theta}_{\rm bkg},\kappa)$ calculated from the 
AMS-02 antiproton data, $p(\boldsymbol{\theta}_{\rm bkg})$ is the prior 
of background parameters $\boldsymbol{\theta}_{\rm bkg}$ which is obtained 
via the MCMC fitting to the B/C, $^{10}$Be/$^9$Be, and proton data, and 
$p(\kappa)$ is the prior of the antineutron-to-antiproton production 
ratio, which is assumed to be Gaussian distribution $N(1.3,0.2^2)$ 
\cite{2003APHHI..17..369F,2014PhRvD..90h5017D}. 

\begin{figure}[!htb]
\includegraphics[width=0.45\textwidth]{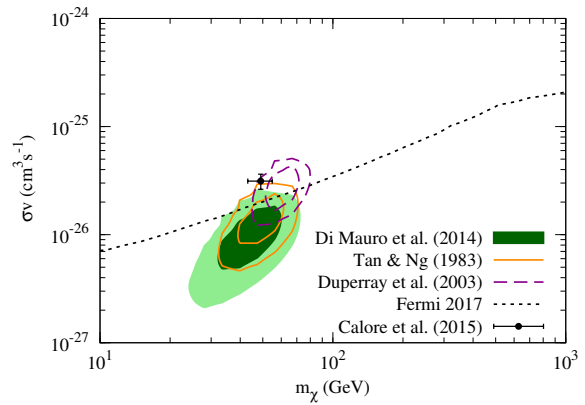}
\caption{Shaded regions and contours are the 68\% and 95\% credible 
regions of parameters $m_{\chi}-\sv$ to fit the antiproton data, for 
three parameterations of the antiproton production cross sections
\cite{1983JPhG....9.1289T,2003PhRvD..68i4017D,2014PhRvD..90h5017D}. 
The annihilation channel is assumed to be $b\bar{b}$. Also shown are 
the Fermi-LAT exclusion limits from observations of dwarf spheroidal 
galaxies \cite{2017ApJ...834..110A}, and the best-fit parameters 
(with a re-scaling of the local density) through fitting to the 
Galactic center GeV excess \cite{2015JCAP...03..038C}.
\label{fig:contour}}
\end{figure}

We find that the AMS-02 data favor a DM component with a mass of a few 
tens GeV and an annihilation cross section of the thermal production level
for quark final state. This conclusion holds for different antiproton 
production cross sections given in Refs. \cite{1983JPhG....9.1289T,
2003PhRvD..68i4017D,2014PhRvD..90h5017D}, as well as different source
distributions of CRs \cite{Sup-C}. The logarithmic Bayes factor 
value ($2\ln K$) of the DM component is found to be about $11-54$ for 
the three cross section parameterizations used. The best fit DM mass is 
about $40-60$ GeV, and the annihilation cross section is about $(1-3)\times 
10^{-26}$ cm$^3$s$^{-1}$ for $b\bar{b}$ channel. Fig. \ref{fig:contour} 
shows the favored parameter regions on the $m_{\chi}-\sv$ plane. 
For DM annihilation into $W^+W^-$, similar results can be found with 
slightly heavier masses (due to the mass threshold to produce $W$ bosons).
Using the PAMELA data, Ref. \cite{2015JCAP...03..021H} obtained similar 
results, although in a suggestive way with significantly larger uncertainties.

It is interesting to note that such a favored parameter region is
consistent with that to fit the GeV $\gamma$-ray excess in the 
Galactic center region \cite{2011PhLB..697..412H,2012PhRvD..86h3511A},
as well as the tentative $\gamma$-ray excesses in the directions of
two dwarf galaxies \cite{2015PhRvL.115h1101G,2016PhRvD..93d3518L}.
Also we find that the favored DM mass is consistent with that inferred
from a tentative $\gamma$-ray line-like signal with energies $\sim43$
GeV from a population of clusters of galaxies \cite{2016PhRvD..93j3525L}.
Such a consistency, if not solely due to coincidence, strongly
supports the common DM origin of the antiproton ``anomaly'' and GeV
$\gamma$-ray excesses. 

\begin{figure}[!htb]
\includegraphics[width=0.45\textwidth]{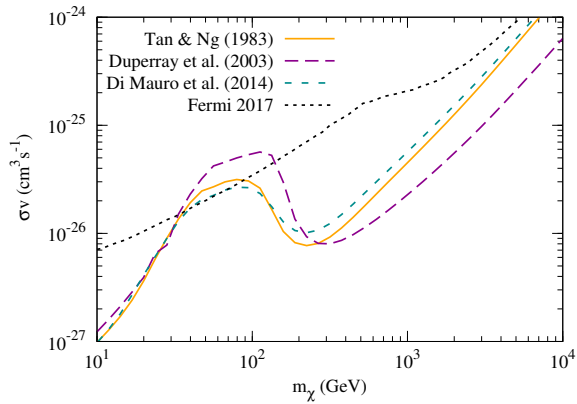}
\caption{The 95\% credible upper limits of the DM annihilation cross 
section versus mass derived through fitting to the AMS-02 data, compared 
with that from Fermi-LAT observations of dwarf spheroidal galaxies 
\cite{2017ApJ...834..110A}.
\label{fig:limit}}
\end{figure}

We also derive the upper limits of the DM annihilation cross section
for DM masses of $10-10^4$ GeV, as shown in Fig. \ref{fig:limit}. Here
the 95\% credible limit of $\sv$ is obtained by setting $\left.
\left(\int_0^{\sv}{\mathcal P}(x){\rm d}x\right)\right/
\left(\int_0^\infty{\mathcal P}(x){\rm d}x\right)=0.95$. Compared with
that derived from the combined analysis of the Fermi-LAT $\gamma$-ray 
emission from a population of dwarf spheroidal galaxies
\cite{2017ApJ...834..110A}, our limits are in general stronger, except 
for the mass range of $30-150$ GeV where we find signal favored by the 
antiproton data. The DM density profiles may affect our constraints by 
a factor of $\lesssim2$, for the Einasto or isothermal profile 
\cite{2015JCAP...09..049J}. On the other hand, the local density 
adopted in this work, 0.3 GeV cm$^{-3}$, may be lower than that from 
recent kinematics measurements \cite{2010A&A...523A..83S}, which makes
our constraints more conservative. 

{\it Conclusion} ---
Compelling evidence indicates that DM particles consist of a substantial 
fraction of the energy density of the Universe. It is also widely 
anticipated that these exotic particles can annihilate with each other 
and produce stable high energy particle pairs, including for example 
electrons/positrons, protons/antiprotons, neutrinos/anti-neutrinos and 
$\gamma$-rays. However, so far no solid evidence for DM annihilation has 
been reported, yet. 

In this work we use the precise measurement of the antiproton flux by
AMS-02 to probe the DM annihilation signal. The CR propagation parameters, 
proton injection parameters, and the solar modulation parameters, which 
are derived through independent fitting to the B/C and $^{10}$Be/$^9$Be 
ratios, and the time-dependent proton fluxes, are taken into account in 
the posterior probability calculation of the DM parameters self-consistently 
within the Bayesian framework. Such an approach does not assume background 
parameters in advance, and thus tends to give less biased results of the 
DM searches.

We find that the antiproton data suggest the existence of a DM signal.
The favored mass of DM particles ranges from 20 to 80 GeV, and the 
annihilation cross section is about $(0.2-5)\times10^{-26}$ cm$^3$s$^{-1}$, 
for $b\bar{b}$ channel. Though further studies are still needed to 
firmly establish the DM origin of the antiproton ``anomaly'', we notice 
that the inferred DM parameters are well consistent with that found in 
the modeling of the Galactic center GeV excess and/or the weak GeV emission 
in the directions of Reticulum 2 and Tucana III. Such a remarkable 
consistency, if not due to coincidence, points towards a common DM 
annihilation origin of these signals. The indication of a similar signal 
from various targets and different messengers will be very important for 
the search for particle DM. For other possibilities to explain the current 
puzzle please see \cite{Sup-D}. We keep in mind that the current framework 
of the CR propagation is relatively simple. More detailed model may be 
necessary for future improvement of the understanding of this problem.

We have obtained the upper limits on the DM annihilation cross section 
from the antiproton data, which are stronger than that set by the 
Fermi-LAT observations of a population of dwarf spheroidal galaxies in 
a wide mass range. The improvement of constraints is expected to be 
benificial from more precise measurements of the data by AMS-02, which 
reduce the uncertainties of both the background and the expectation of 
the signal. Our improved method also helps because the background 
parameters are taken into account with proper likelihood instead of a 
choice by hand. 

{\it Note:} --- Recently, Ref. \cite{2016arXiv161003071C} appears on arXiv. 
We have different approaches but consistent results.

\acknowledgments

This work is supported in part by the National Basic Research Program of 
China (No. 2013CB837000), the National Key Research and Development Program 
of China (No. 2016YFA0400200), the National Natural Science Foundation 
of China (No. 11525313), and the 100 Talents program of Chinese Academy 
of Sciences.

\bibliographystyle{apsrev}
\bibliography{refs}

\include{pbar-sup}

\end{document}

%% file: pbar-sup.tex
\def\nat{Nature\ }
\def\aap{Astron.\ Astrophys.\ }
\def\apj{Astrophys.\ J.\ }
\def\apjl{Astrophys.\ J.\ Lett.\ }
\def\apjs{Astrophys.\ J.\ Supp.\ }
\def\aj{Astron.\ J.\ }
\def\mnras{Mon.\ Not.\ Roy.\ Astron.\ Soc.\ }
\def\physrep{Phys.\ Rept.\ }
\def\prd{Phys.\ Rev.\ D\ }
\def\prl{Phys.\ Rev.\ Lett.\ }
\def\apss{Astrophys.\ Space\ Sci.\ }
\def\araa{Annu.\ Rev.\ Astron.\ Astrophys.\ }
\def\jcap{J.\ Cosmol.\ Astropart.\ Phys.\ }

\def\sv{\ensuremath{\langle\sigma v\rangle}}



\section*{Supplemental Material}

\subsection{Background parameters}

The background parameters used to calculate the secondary antiproton
fluxes include the propagation, source spectra, and solar modulation 
parameters. The major propagation parameters are: the rigidity-dependent
diffusion coefficient $D(R)=\beta D_0(R/4\,{\rm GV})^{\delta}$ where
$R$ is the rigidity of a particle and $\beta$ is the velocity in unit 
of speed of light, the half-height of the propagation halo $z_h$, and
the Alfvenic speed $v_A$ which characterizes the reacceleration. The
injection spectrum of CR nuclei is assumed to be a broken power-law
in rigidity, with indices $\nu_1$ and $\nu_2$ below/above the break 
rigidity $R_{br}$. The post-propagated proton flux is normalized to
$A_p$. The spatial distribution of CR sources is parameterized as
$$f(r,z)=(r/r_s)^{1.25}\exp\left[-3.56(r-r_s)\right]
\exp(-|z|/z_s),$$
where $r_s=8.5$ kpc and $z_s=0.2$ kpc. Such a distribution follows the 
pulsar distribution with parameters adjusted slightly based on the 
diffuse $\gamma$-ray data [22]. Finally, there are 
two solar modulation parameters, $\Phi_0$ and $\Phi_1$, as described in 
the main text. The parameters values obtained in the MCMC fittings of 
Ref. [26] are given in Table \ref{table:para}.

\begin{table}[!htb]
\caption{Mean values and posterior $1\sigma$ uncertainties of the parameters 
for the fitting with B/C, $^{10}$Be/$^9$Be, and the proton spectra.}
\begin{tabular}{cccccccc}
\hline \hline
 Parameter & Unit & Value\\
\hline
$D_0$ & ($10^{28}$cm$^2$ s$^{-1}$) & $7.24\pm0.97$ \\
$\delta$ & & $0.380\pm0.007$ \\
$z_h$ & (kpc) & $5.93\pm1.13$ \\
$v_A$ & (km s$^{-1}$) & $38.5\pm1.3$ \\
$\log(A_p^a)$ & & $-8.347\pm0.002$ \\
$\nu_1$ & & $1.69\pm0.02$ \\
$\nu_2$ & & $2.37\pm0.01$ \\
$\log(R_{br}^b)$ & & $4.11\pm0.02$ \\
$\Phi_0$ & (MV) & $180\pm8$ \\
$\Phi_1$ & (MV) & $487\pm11$ \\
\hline
\end{tabular}\\
Notes: $^a$Propagated flux normalization of protons at $100$ GeV in 
unit of cm$^{-2}$s$^{-1}$sr$^{-1}$MeV$^{-1}$.
$^b$Break rigidity of the proton injection spectrum in unit of MV.
\label{table:para}
\end{table}

\subsection{Hadronic $pp$ interactions}

\begin{figure*}[!htb]
\includegraphics[width=0.45\textwidth]{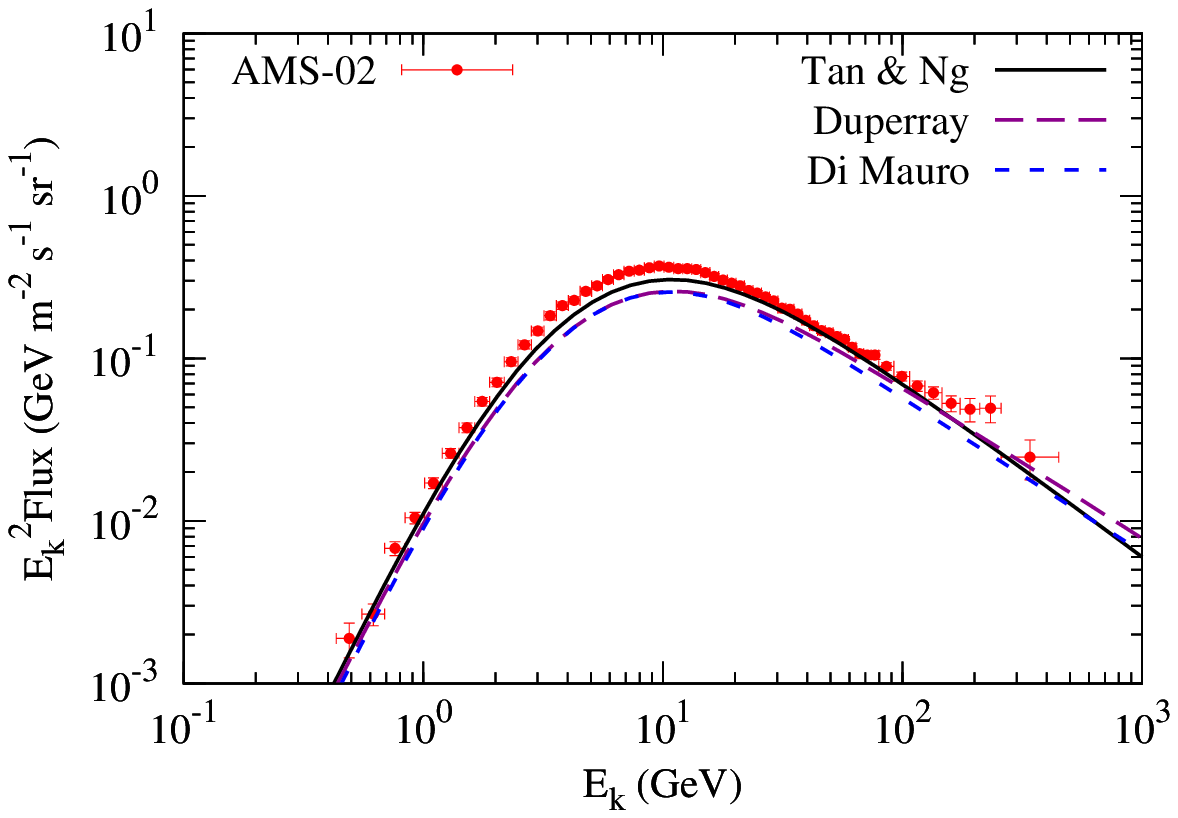}
\includegraphics[width=0.45\textwidth]{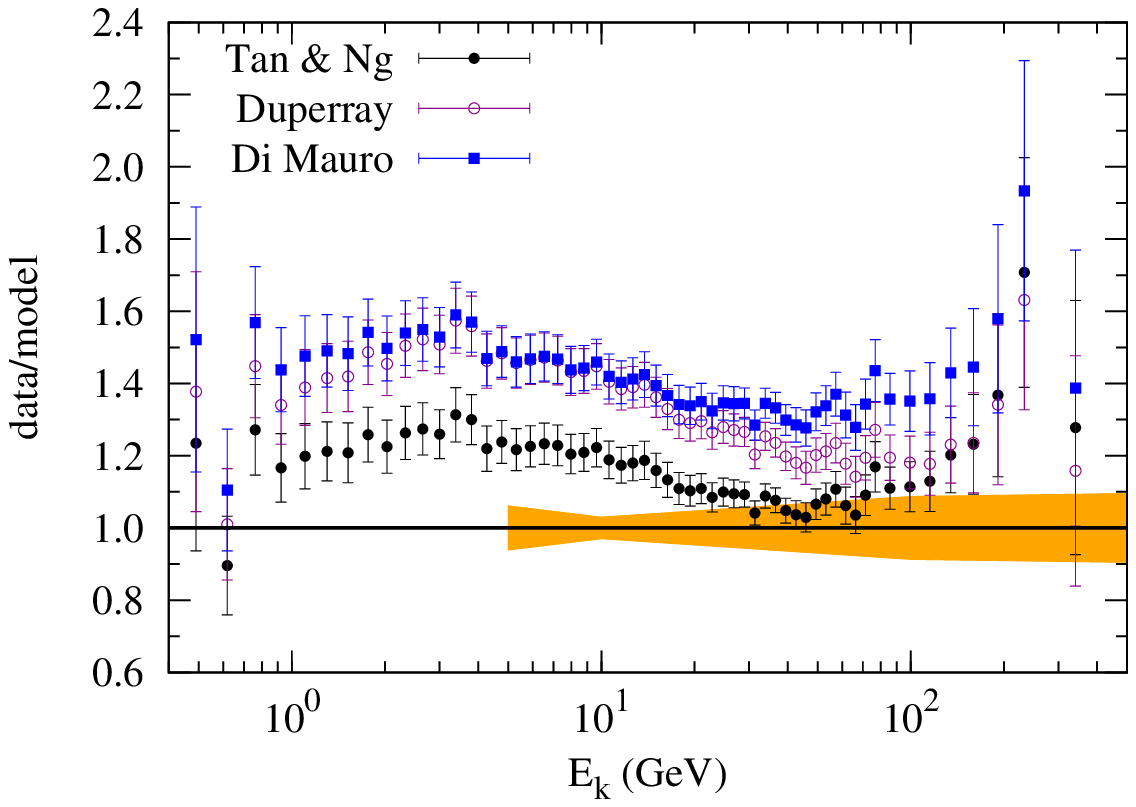}
\caption{Left: antiproton fluxes for different parameterizations of 
$pp$ collision induced antiproton yield spectra. Right: ratios of data 
to model predictions for different $pp$ interaction parameterizations. 
The shaded region shows the uncertainty bands of the parameterization given 
in Ref. [39]. In this plot $\kappa=1$ is assumed.
\label{fig:test_int}}
\end{figure*}

There are relatively large uncertainties of the hadronic interaction
models of inelastic $pp$ collisions. The default model of this work
uses the parameterization given by Tan \& Ng [37]. With more and more 
experimental data, updated parameterizations were available in literature 
[38-40]. Note further that there are heavy nuclei in both 
the CRs and the ISM. A nuclear enhancement factor from Monte Carlo 
simulations, $\epsilon=1.58\times(E_k/{\rm GeV})^{0.034}$, is employed 
[42]. In the left panel of Fig. \ref{fig:test_int} 
we show the comparison of results for several different parameterizations 
of antiproton production from $pp$-collision. We find that there are
several tens percents difference between each other. However, the
qualitative conclusion that there are low energy deficits of antiprotons
compared with the AMS-02 data keeps unchanged. The logarithmic Bayes 
factor of adding a DM component to the background model, is about
19 for cross sections of Ref. [37], 54 for Ref. [38], and 11 for Ref.
[39]. Note that we use Eq. (12) in Ref. [39] in this work. Another 
formula (Eq. (13)) was also given in Ref. [39]. Using Eq. (13) we 
find that the background antiproton flux is overall lower by a factor 
of $\sim1.35$ than the data, but no significant DM signal is shown.
The problem is that Eq. (13) may give unrealistic extrapolation of 
the results out of the energy coverage by the data which is about 
$4-500$ GeV, due to a cubic term of the $p_T$-dependence [39].
Also the spline interpolation method in Ref. [39] gives similar result 
as that of Eq. (12) in the energy range we are mostly interested in 
(GeV to tens of GeV).

The right panel of Fig. \ref{fig:test_int} shows ratios of data to model 
predictions of antiproton fluxes for the three $pp$ interaction 
parameterizations. It is clear to show that there are all low energy 
excesses of the data compared with the predictions, even when a constant
rescaling factor $\kappa$ is applied to the models. In particular, the 
excesses can not be simply ascribed to the uncertainties of the 
parameterizations, as illustrated by the shaded region for that of Ref. [39].

\subsection{Test of CR source distributions}

\begin{figure*}[!htb]
\includegraphics[width=0.45\textwidth]{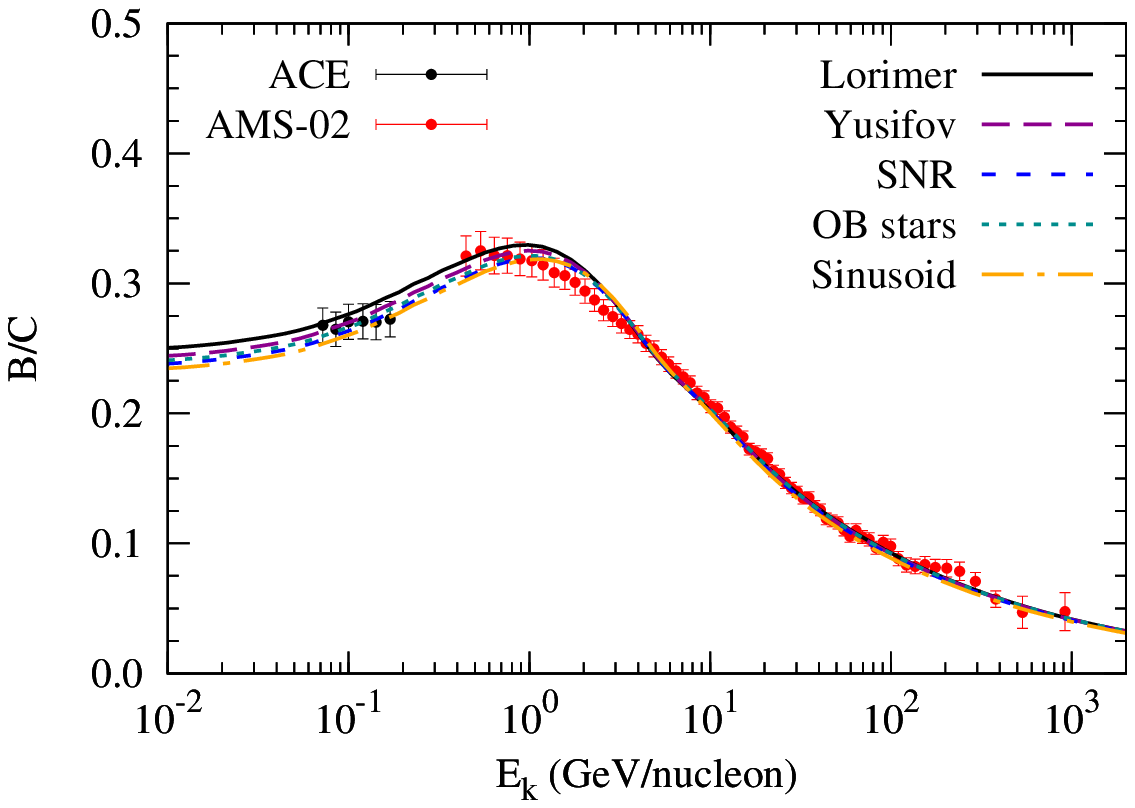}
\includegraphics[width=0.45\textwidth]{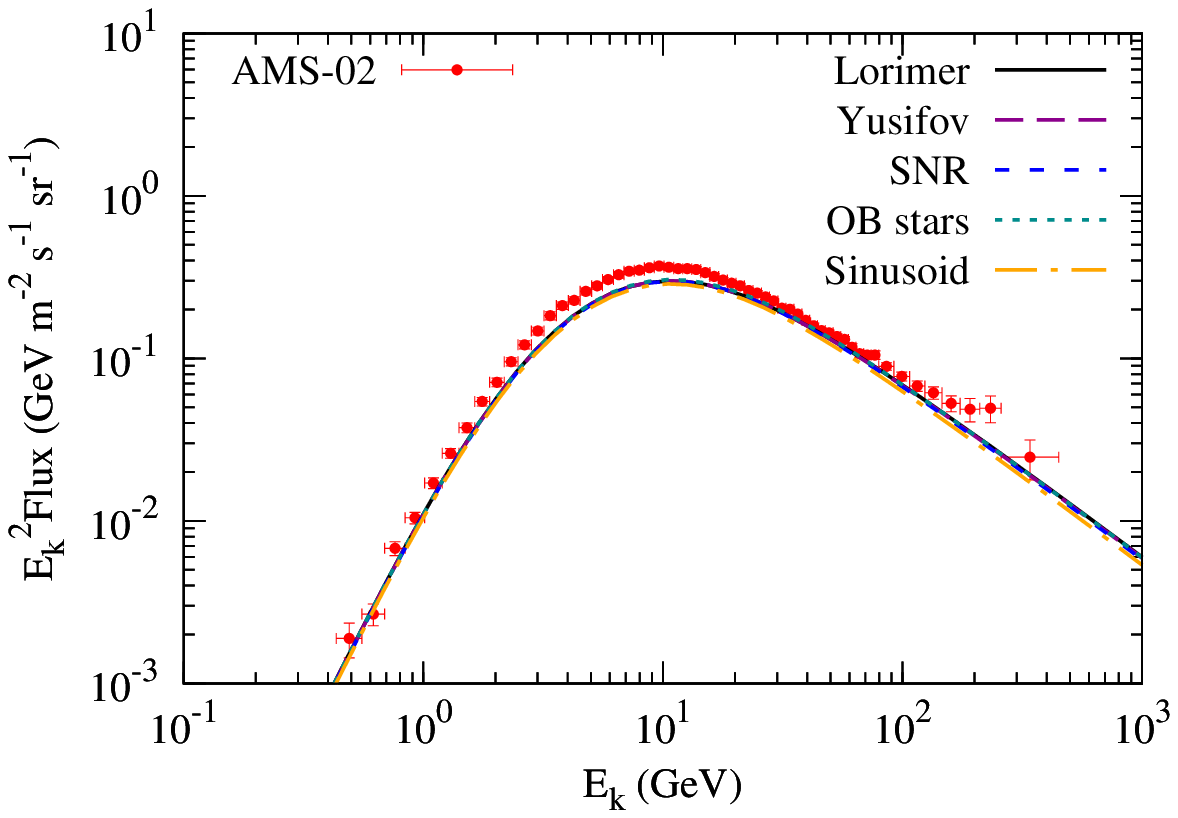}
\caption{The B/C ratios (left) and antiproton fluxes (right) for different
source distributions: pulsars from Lorimer et al. [48], pulsars from Yusifov 
\& Kucuk [49], supernova remnants [50], OB stars [51], and a sinusoid function.
\label{fig:test_src}}
\end{figure*}

The origin of CRs is still in debate, and hence the source distributions
of CRs are essentially unknown. Here we test the effect of different source 
distributions. We consider the four kinds of radial distributions which
were discussed in Ref. [47] to fit the diffuse $\gamma$-ray data of Fermi-LAT. 
They are: two parameterizations of pulsar distributions [48,49], supernova 
remnant (SNR) distribution [50], and the massive OB star distribution [51]. 
Also we know that there are
spiral structures of stars and gas in the Galaxy. We thus employ a 
sinusoidal radial distribution function, $f(r)=1+\sin(r/0.67\,{\rm kpc})$,
to mimic the spiral arm-like distribution of CR sources. The $z$-direction
distribution is adopted to be $\propto\exp(-|z|/0.2\,{\rm kpc})$ for all 
source functions. For each source distribution, the background parameters 
are identical to the mean values given in Table \ref{table:para}, except 
for $D_0$ which is slightly adjusted to match the B/C data. As shown in 
Fig. \ref{fig:test_src}, different source distributions lead to very 
little difference of both the B/C ratios and antiproton fluxes.

\begin{figure*}[!htb]
\includegraphics[width=0.45\textwidth]{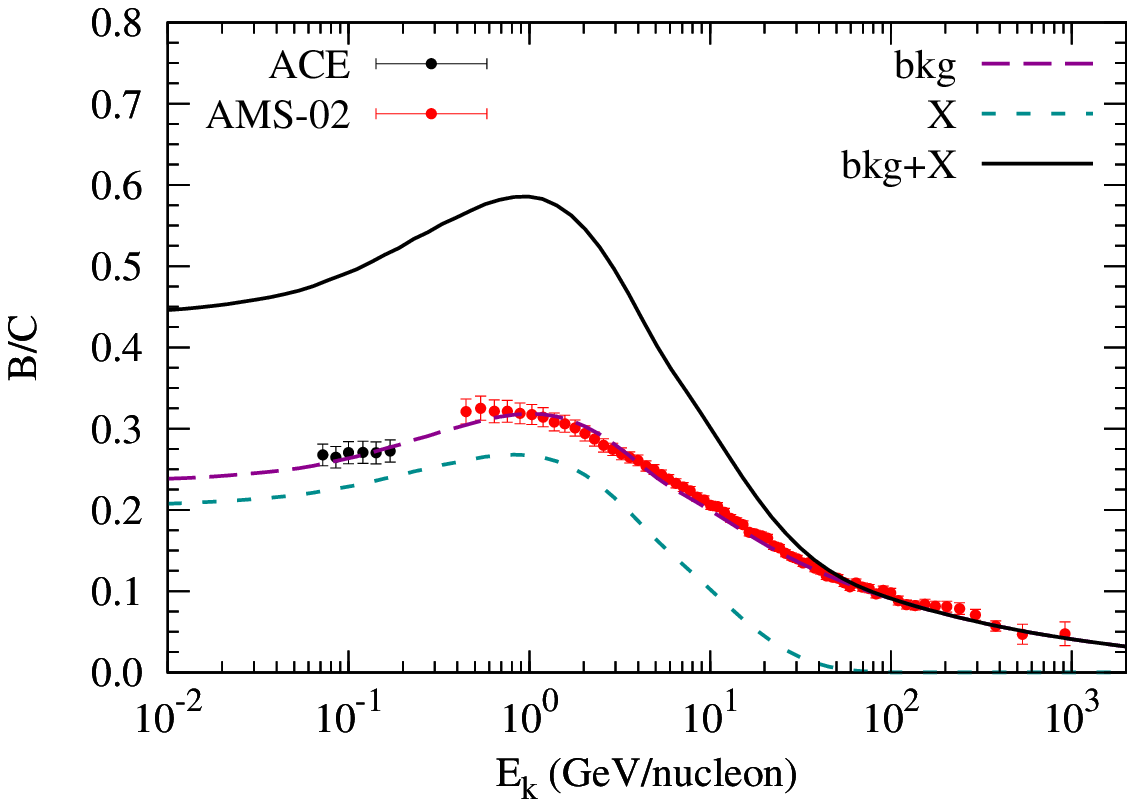}
\includegraphics[width=0.45\textwidth]{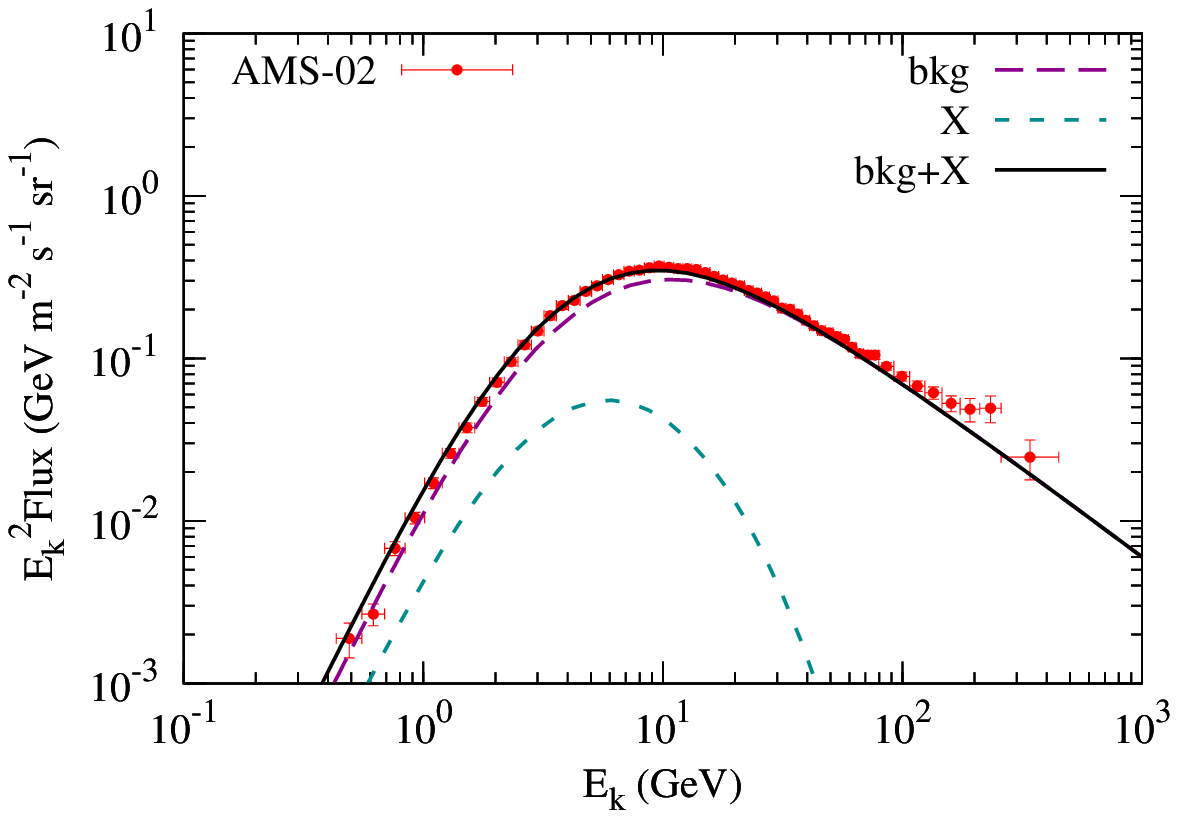}
\caption{The B/C ratio (left) and antiproton flux (right) for the
model with an astrophysical component ``$X$''. 
\label{fig:test_ast}}
\end{figure*}

\subsection{Astrophysical interpretations}

It is interesting to investigate whether there are astrophysical sources 
other than the DM annihilation, which can produce the excess antiprotons 
without violating the B/C data. Generally speaking, an astrophysical 
source (for instance, SNR) will produce secondary antiprotons and nuclei
simultaneously. We consider such an ``$X$'' component, assuming the same 
source parameters as the background CRs except that an exponential cutoff 
with characteristic rigidity of 30 GV is assumed, in order to fit the 
antiproton data. Also we assume that this ``$X$'' component contributes 
negligible primary CRs. A potential astrophysical realization is the 
acceleration and interaction of CRs within molecular clouds [61].
Fig. \ref{fig:test_ast} shows the results 
for the scenario with such an ``$X$'' component. We find that to produce
proper amount of antiprotons, the model over-produces Boron nuclei by
a factor of $\sim2$. Only if the relative abundance of Carbon/Oxygen
to protons is a factor of about 30 lower than that of the background
sources, the model can simultaneously acount for the antiproton and B/C 
data. Note that the ``$X$-to-bkg'' ratio of antiprotons is smaller 
than that of the B/C ratio. This is because antiprotons are produced by 
parent protons with much higher energies which are suppressed due to the 
cutoff of primary CRs for the ``$X$'' component. We have tested that
when the cutoff rigidity approaches infinity, the ``$X$-to-bkg'' ratio
of antiprotons asymptotically approaches that of the B/C.

An alternative astrophysical scenario to reconcile the antiproton and 
B/C data was proposed in Ref. [62], in which a local and fresh source 
(perhaps associated with the local bubble) is postulated to contribute 
additional low energy primaries and decrease the measured 
secondary/primary ratio. Since the local source will also produce
protons, the injection spectrum of the Galactic component of protons
need to be significantly suppressed below several GeV. This may not
affect the antiproton production due to a high threshold energy, but
may imprint in the diffuse $\gamma$-rays. Further studies of the
consistency between the model prediction and the Fermi-LAT diffuse 
$\gamma$-ray data may be interesting to test the model.

In Ref. [63], the authors suggested an empirical adjustment of the 
diffusion coefficient with a velocity-dependent term $\beta^{\eta}$ to 
explain the B/C and antiproton data. However, the physical reason for 
such an adjustment is not well justified. 

We suggest that the DM annihilation scenario is a simple model that can 
reasonably reproduce the antiproton data. The inferred physical parameters
are in agreement with several other hints from $\gamma$-ray observations 
of various targets. Therefore, it may be the time to seriously consider
the possibility of their DM origin.

